\documentclass[article,12pt,nofootinbib]{revtex4}

\usepackage[paper=letterpaper,margin=1.5in]{geometry}

\usepackage{epsfig,color}
\usepackage{graphicx}
\usepackage{amssymb,amsmath}
\usepackage{time}
\usepackage[varg]{txfonts}
\usepackage{hyperref}
\hypersetup{
  colorlinks,
  citecolor=green,
  linkcolor=blue}
\newcommand{\be}{\begin{equation}}
\newcommand{\ee}{\end{equation}}
\newcommand{\bc}{\begin{center}}
\newcommand{\ec}{\end{center}}

\DeclareMathOperator\erfc{erfc}

\renewcommand{\vec}[1]{\textnormal{\boldmath$#1$}}
\begin{document}

\bibliographystyle{revtex}

\begin{flushright}
{\normalsize
DESY 16-138\\
July 2016}
\end{flushright}

\vspace{.4cm}

\title
{Corrugated structure insertion for extending the SASE bandwidth up to 3\% at the European XFEL}
\author{I. Zagorodnov, G. Feng, T. Limberg}
\affiliation{Deutsches Elektronen-Synchrotron, Notkestrasse 85,
22603 Hamburg, Germany}
\date{\today}

\vspace{.4cm} 
\begin{abstract} 
The usage of x-ray free electron laser (XFEL) in femtosecond nanocrystallography involves sequential illumination of many small crystals of arbitrary orientation. Hence a wide radiation bandwidth will be useful in order to obtain and to index a larger number of Bragg peaks used for determination of the crystal orientation.

Considering the baseline configuration of the European XFEL in Hamburg, and based on beam dynamics simulations, we demonstrate here that the usage of corrugated structures allows for a considerable increase in radiation bandwidth. Data collection with a 3\% bandwidth, a few microjoule radiation pulse energy, a few femtosecond pulse duration, and a photon energy of 5.4 keV is possible.
 
 For this study we have developed an analytical {\it modal} representation of the  short-range wake function of the flat corrugated structures for arbitrary offsets of the source and the witness particles.

{\it Keywords}: wakefields, impedance, corrugated structure, longitudinal phase space control, free electron laser, bandwidth.  
\end{abstract}

\maketitle

%
\section{Introduction}

X-ray crystallography is a method for imaging molecules with atomic resolution. The usage of x-ray free electron laser (XFEL) in femtosecond nanocrystallography involves sequential illumination of many small crystals of arbitrary orientation~\cite{Chapman}.  The success of nanocrystallography depends on the robustness of the procedure for pattern determination. After acquisition, diffraction patterns are analyzed to assign indexes to Bragg peaks. Indexing algorithms used in crystallography are used to determine the orientation of the diffraction data from a single crystal when a relatively large number of reflections are recorded. The number of Bragg peaks is proportional to the bandwidth of the incident radiation pulse. With nominal scenario for beam compression and transport at the European XFEL~\cite{EXFEL} the radiation bandwidth is quite narrow~\cite{Yurkov2011}, on the level of several permilles. 

It was shown in~\cite{Serkez2013} that a strong beam compression could be used to increase the correlated energy spread of the electron beam and as consequence to increase the radiation bandwidth up to 2 \% at photon energy  6 keV. In this paper we study another possibility: we use over-compression of the electron beam at the last chicane and insert corrugated structures before the undulator section. These structures are used to generate strong wakefields which introduce a large energy drop along the bunch~\cite{Bane2012}.

Considering the baseline configuration of the European XFEL, and based on beam dynamics simulations, we demonstrate here that the usage of corrugated structures allows for a tenfold increase in radiation bandwidth. Data collection with a 3\% bandwidth, a few microjoule radiation pulse energy, a few femtosecond pulse duration, and a photon energy of 5.4 keV is possible. 

For this study we have developed an analytical {\it modal} representation of the  short-range wake function of the flat corrugated structures for arbitrary offsets of the source and the witness particles. This representation is different from ones published so far,~\cite{Bane2003}-\cite{Bane2016b},  and agrees well with the numerical results obtained by direct numerical solution of Maxwell's equations for electron bunch shapes of our interest.

%
\section{Beam dynamics in linear accelerator of the European XFEL}\label{sec:2}

\begin{figure}[htbp]
\centering
\includegraphics*[height=50mm]{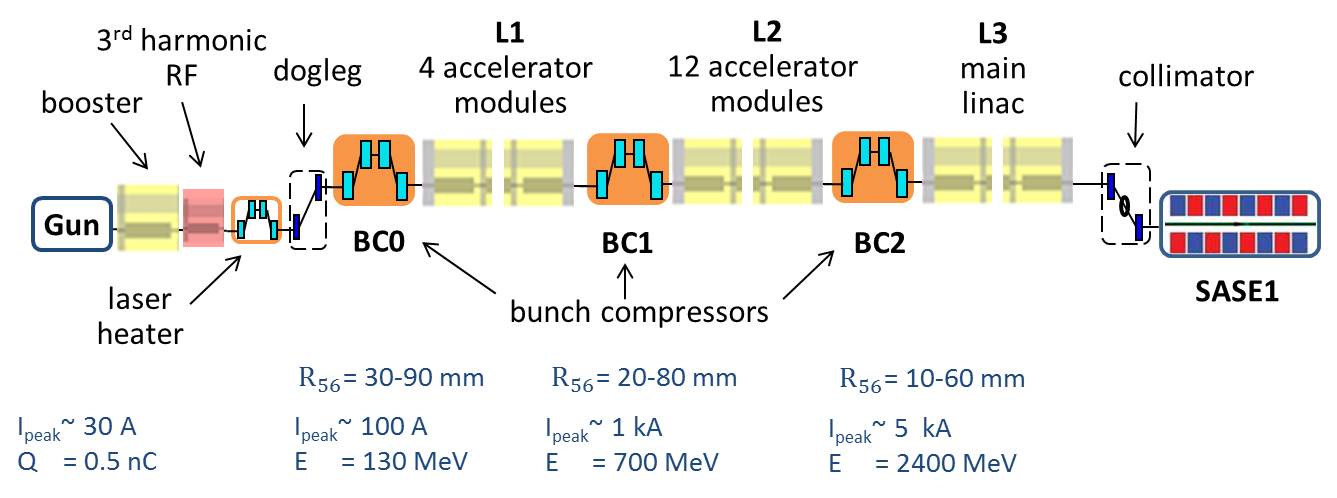}
\caption{Layout of the European XFEL.}\label{Fig01}
\end{figure}

\begin{figure}[htbp]
	\centering
	\includegraphics*[height=50mm]{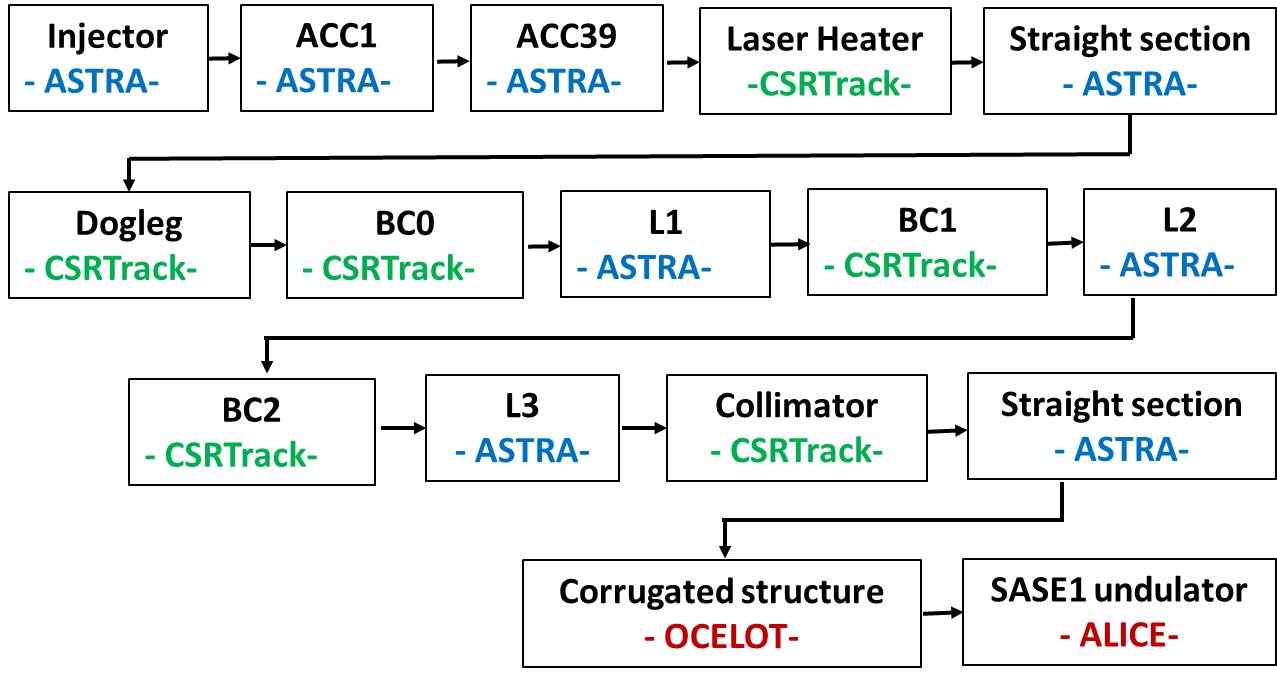}
	\caption{Setup of the beam dynamics simulations of the European XFEL.}\label{Fig02}
\end{figure}

The European XFEL is under construction now in Hamburg. It contains several photon lines to produce extremely short and bright light pulses with wave length down to 0.04 nm. The layout of the accelerator (with the undulator line SASE1) is shown in Fig.~\ref{Fig01}. The linear accelerator (linac) contains many radio-frequency (RF) accelerating modules, 3rd harmonic RF module for longitudinal phase space linearization, three bunch compressors etc. Beam dynamics in the accelerator with nominal set of parameters is described, for example, in~\cite{Feng2013}. 

In this paper we consider a special scenario with over-compression of the electron bunch in the last bunch compressor (BC2 in Fig.~\ref{Fig01}). The over-compression allows to obtain a bunch with lower particle energy at the tail. Hence the wake fields produced by linac elements following the chicane BC2 will not compensate (as it appears in the nominal scenario's, see ~\cite{Feng2013}), but increase the energy deviation between the particles at the head and at the tail of the electron bunch. This energy difference will be enhanced later on by strong wake fields of the corrugated structures as described in Section~\ref{sec:4}.

\begin{table}[htbp]
	\centering
	\caption{Parameters of the linac at the simulation.}
	\label{Table01}
	\begin{tabular}{lcc}
		{\bf Parameter name}					&{\bf Value}	&{\bf Unit}\\
		Voltage of booster, $V_1$   				& 157.31 		& MV \\
		Phase of booster, $\phi_1$  				& 21.76			& deg \\
		Voltage of third harmonic module, $V_{13}$  	& 24.76 	& MV \\
		Phase of third harmonic module, $\phi_{13}$ 	& 203.82	& deg \\
		Momentum compaction in BC0, $R_{56}$  	& 54.8			& mm \\
		Voltage of linac0, $V_2$   				& 673.54 		& MV \\
		Phase of linac0, $\phi_2$  				& 32.16			& deg \\
		Momentum compaction in BC1, $R_{56}$  	& 50			& mm \\
		Voltage of linac1, $V_3$   				& 1776.25 		& MV \\
		Phase of linac1, $\phi_3$  				& 16.75			& deg \\
		Momentum compaction in BC2, $R_{56}$  	& 28.3			& mm \\
		Final beam energy, $E$  				& 14			& GeV \\
	\end{tabular}
\end{table}

The setup of simulations is sketched in Fig.~\ref{Fig02}. For the numerical modeling of the linac we have used two codes: code ASTRA~\cite{ASTRA} for straight sections and code CSRtrack~\cite{CSRtrack} for curved parts of the beam trajectory. Code ASTRA allows to take into account three dimensional space-charge fields, but it neglects the radiation. Code CSRtrack is used to model coherent synchrotron radiation in the dispersion sections. Both codes neglect the impact of the vacuum chamber on the beam (code CSRtrack can model boundary as two perfectly conducting parallel plates, but we have not used this boundary approximation in the simulations). In order to take effect of the vacuum chamber into account we have applied the concept of wake fileds ~\cite{Zotter}. The wake functions for a point charge at different linac components are obtained with the code ECHO~\cite{ECHO},~\cite{RFWakes}. The impact of wake fields has been modeled through discrete kicks at several positions along the linac.

In order to find the RF parameters of accelerating modules we have used the approach introduced  in~\cite{Zag2011}. The parameters obtained with this method are listed in Table~\ref{Table01}. In the following we describe properties of the particle distribution at the end of the linac from beam dynamics simulations with these RF settings.

\begin{figure}[htbp]
	\centering
	\includegraphics*[height=60mm]{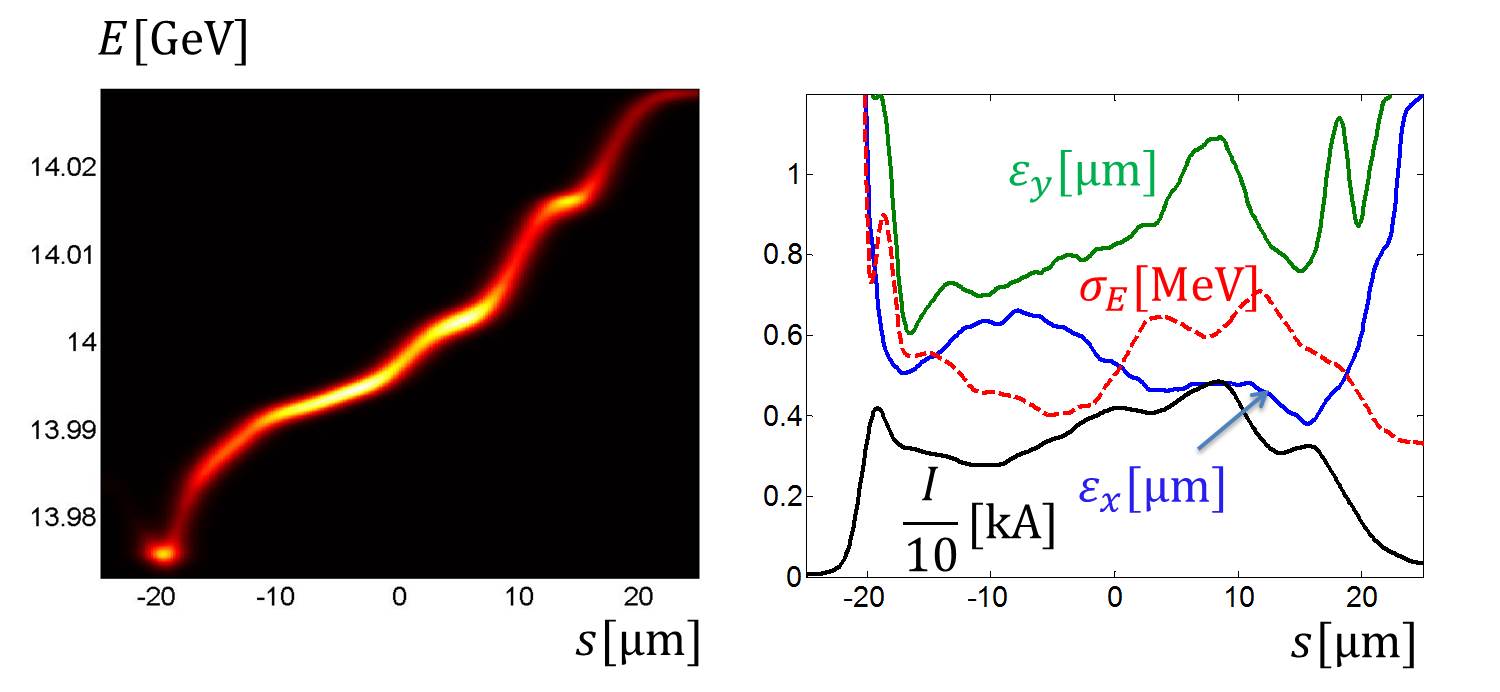}
	\caption{Longutudinal phase space of the electron bunch after collimator (left plot). Slice parameters of the same beam (right plot).}\label{Fig03}
\end{figure}

 The longitudinal phase space and the slice parameters of the beam after collimator section are shown in Fig.~\ref{Fig03}. The projected normalized emittance in $x$-plane is equal to 0.72 $\mu$m, in $y$-plane - 1.17 $\mu$m. The parameter of the main interest for us is the difference in the energy between the beam head and the beam tail. For the beam overcompressed in the last bunch compressor we obtain that the tail has an energy drop of about 60 MeV. 

 It is well known~\cite{Saldin},  that the spectrum bandwidth of self-amplified spontaneous emission (SASE) radiation is approximately equal to the doubled energy bandwidth in the electron beam. For the final beam energy of 14 GeV the relative energy bandwidth in the electron beam is about 0.4\% and we can hope to have the SASE radiation bandwidth of 0.8\%. In order to increase the bandwidth up to several percents we will use a chain of corrugated structures. 
 
 In our studies the corrugate structure has the parameters suggested at SLAC~\cite{Zhang2015}.  The appropriateness of this choice of parameters for LCLS applications is partially addressed in Sec. III of~\cite{Zhang2015}. The European XFEL has the electron beam parameters which are close to the beam parameters of LCLS and the same arguments hold.
 
 In the next section we describe the calculation of wakefields of the corrugated structure and give an analytical expression for the wake function of a point charge. The obtained analytical representation will be used in beam dynamics simulations presented in Section~\ref{sec:4}.

%
\section{Analytical representation of wake function of flat corrugated structure }\label{sec:3}

\begin{figure}[htbp]
  	\centering
  	\includegraphics*[height=50mm]{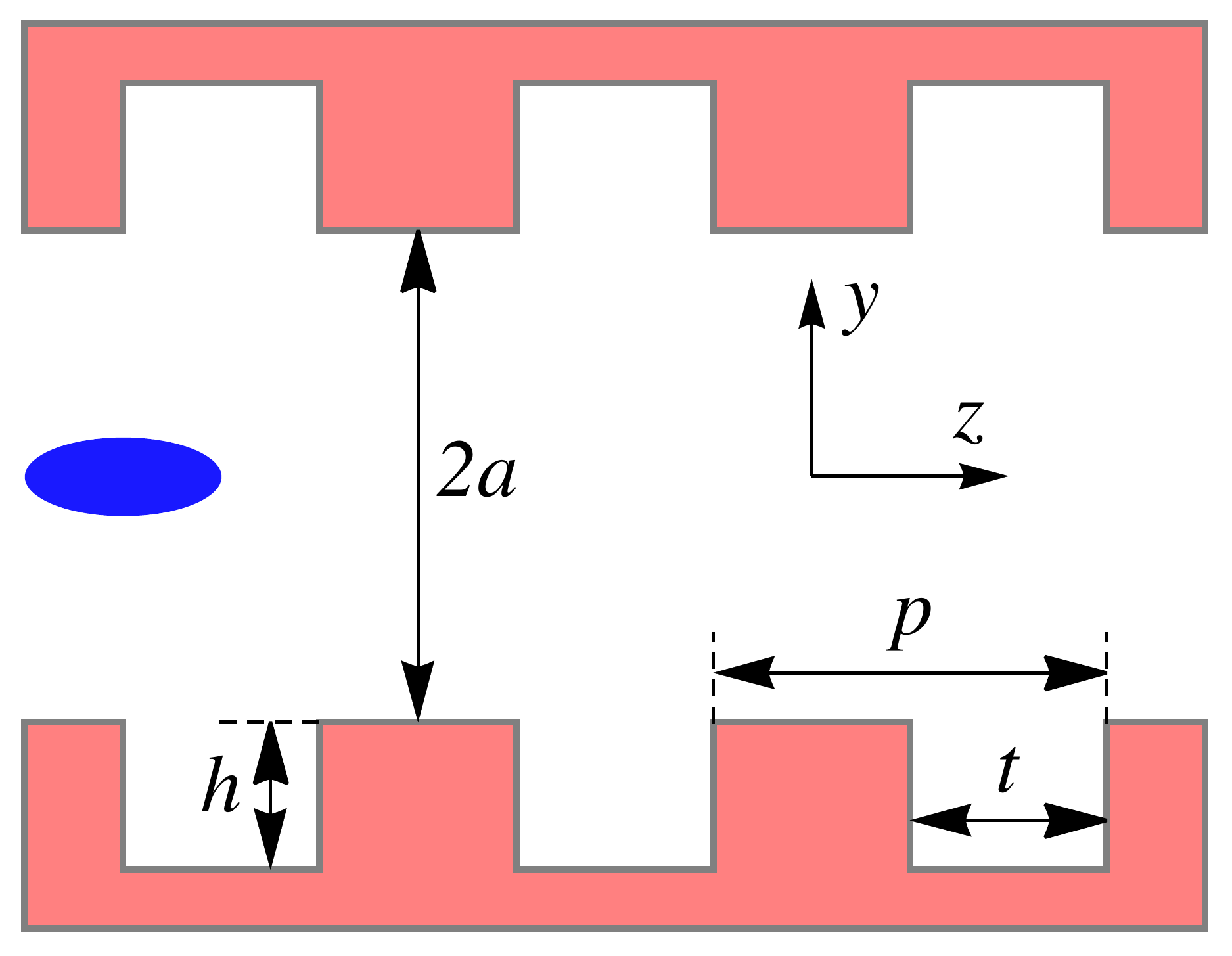}
  	\caption{Gerometry of the corrugated structure. The blue ellipse represents an electron
  		beam propagating along the z axis.}\label{Fig04}
  \end{figure}

 The geometry of the corrugated structure is shown in Fig.~\ref{Fig04}.  It is horizontally oriented, of full gap $2a$, with symmetry planes at $y=0$ and $x=w$ . The values of period $p$, corrugation depth $h$, longitudinal gap $t$ and plate halfwidth $w$ (in $x$-direction) are listed in  Table~\ref{Table02}. 
 
 There are several publications with different analytical expressions for the long-range and the short-range wake functions of such structure, ~\cite{Bane2003}-\cite{Bane2016b}. In this paper we are interested only in the short-range three dimensional wake function for arbitrary position of the source and the witness particles.  In~\cite{Novo} an estimation of the longitudinal wake function on axis was obtained from numerical electromagnetic calculations. In \cite{Bane2016a} the short range limit of three dimensional wake function was derived analytically. Later, in \cite{Bane2016b}, this "zeroth"-order approximation was improved and extended to "first"-order model. In this paper we will present an alternative {\it modal} expression which agrees asymptotically with analytical representations of~\cite{Bane2016b}.
  
  \begin{table}[htbp]
  	\centering
  	\caption{Parameters of the corrugated structure.}
  	\label{Table02}
  	\begin{tabular}{lcc}
  		{\bf Parameter name}					&{\bf Value}	&{\bf Unit}\\
  		Depth, $h$	   							& 0.5 			& mm \\
  		Gap, $t$  								& 0.25			& mm \\
  		Period, $p$  							& 0.5 			& mm \\
  		Half aperture, $a$ 						& 0.7			& mm \\
  		Half width, $w$  						& 0.6			& mm \\
  		Length, $L$   							& 2 			& m \\
  	\end{tabular}
  \end{table}

The corrugated structure belongs to the class of structures having rectangular cross-section, where the height can vary as function of longitudinal coordinate but the width and side walls remain fixed. For such structures, a
Fourier representation of the wake potential through one-dimensional functions has been derived in \cite{Zag2015}. It follows from the harmonicy of the wake potential in source, $x_0, y_0$, and witness, $x, y$,  transverse coordinates that the longitudinal coupling impedance of the rectangular structure of arbitrary width $2w$ can be written as
\begin{align}\label{Eq01}
Z_z(x_0,y_0,x,y,k)=\frac{1}{w}\sum\limits_{m=1}^\infty 
Z_z(y_0,y,k_{x,m},k)\sin(k_{x,m} x_0)\sin(k_{x,m} x),\\ 
0<x<2w,\quad -a<y<a,\quad k_{x,m}=\frac{\pi}{2w}m,\quad k=\frac{2\pi}{\lambda},\nonumber
\end{align}
where the modal impedance $Z_z(y_0,y,k_x,k)$ is the same as the one for a flat structure infinite in $x$-direction, $\lambda$ is the wavelength. From the geometrical symmetries of the flat structure the following representation of the modal impedance could be derived
\begin{equation}\label{Eq02}
Z_z(y_0,y,k_x,k)=Z_z^{cc}(k_x,k)\cosh(k_x y_0)\cosh(k_x y)+Z_z^{ss}(k_x,k)\sinh(k_x y_0)\sinh(k_x y).
\end{equation}

Let us suggest that  the beam-boundary interaction can be described by a surface impedance $\eta(k)$~\cite{Bane2015a} . It was shown in~\cite{Bane2015a} that the  modal impedance coefficients for the infinite flat structure can be presented as
\begin{align}\label{Eq03}
&Z_z^{cc}(k_x,k)=\frac{Z_0 c}{2a \cosh^2(X)} \left[\eta(k)^{-1}-ika\frac{\tanh(X)}{X}\right]^{-1}\nonumber\\
&Z_z^{ss}(k_x,k)=\frac{Z_0 c}{2a \sinh^2(X)} \left[\eta(k)^{-1}-ika\frac{\coth(X)}{X}\right]^{-1},\qquad X=a k_x.
\end{align}

The high frequency limit, $\eta(k)^{-1}<<k$, of the impedance was considered in~\cite{Bane2016a}. At this limit the both modal coefficients, Eq.~\ref{Eq03}, reduce to the same expression
\begin{equation}\label{Eq04}
Z_z^{cc}(k_x,k)=Z_z^{ss}(k_x,k)=i\frac{Z_0 c}{2ka^2 \cosh(X)\sinh(X)}.
\end{equation}

This "zeroth"-order approximation gives for sufficiently short bunches 
a reasonable approximations of the wakefields. In~\cite{Bane2016a} it was shown that, when
the full bunch length is 30 $\mu m$, the zeroth order (on-axis) wakes agree to within
20-30\% with numerical ECHO~\cite{Zag2015} results. In~\cite{Bane2016b} a "first"-order approximation was introduced. In this paper we follow the same approach as in~\cite{Bane2016b}, but we use this "first"-order approximation not in total but in {\it modal} representation of the wake function. 

It was shown in~\cite{Bane2016b} that the interaction of the charged beam with the perfectly conducting flat corrugated structure can be described by the same surface impedance as for an infinite chain of pillboxes~\cite{Bane1999}:
\begin{align}\label{Eq05}
\eta(k)=\frac{1+i}{\sqrt{k}}\left [\alpha\left(\frac{t}{p}\right) L \sqrt{\frac{\pi}{t}} \right]^{-1},\qquad
\alpha(x)=1-0.465\sqrt{x}-0.070x.
\end{align}

The modal wake function in time domain can be obtained from modal impedance in frequency domain (Eq.~(\ref{Eq03}),~Eq.~(\ref{Eq05})) by inverse Fourier transform. The exact result of the Fourier transform can be found at the same way as in ~\cite{Bane1999}
\begin{align}\label{Eq05b}
w_z^{cc}(k_x,s)=\frac{Z_0 c X}{a\sinh(2X)}e^{\frac{s}{s_0^{*}}}\erfc\left(\sqrt{\frac{s}{s_0^{*}}}\right),\quad s_0^{*}=\frac{\tanh^2(X)}{X^2}\frac{16}{\pi}s_0,\nonumber\\ 
w_z^{ss}(k_x,s)=\frac{Z_0 c X}{a\sinh(2X)}e^{\frac{s}{s_0^{**}}}\erfc\left(\sqrt{\frac{s}{s_0^{**}}}\right),\quad s_0^{**}=\frac{\coth^2(X)}{X^2}\frac{16}{\pi}s_0,
\end{align}
with the distance scale factor
\begin{equation}\label{Eq06}
s_0=\frac{t}{8}\left[\frac{a}{\alpha p}\right]^2.
\end{equation}
Here $s$ is the distance in the longitudinal coordinate between the source and the witness particles.
In the following we will replace the exact result, Eq.~(\ref{Eq05b}), by an exponential approximation  
\begin{align}\label{Eq07}
w_z^{cc}(k_x,s)=\frac{Z_0 c X}{a\sinh(2X)}e^{-\frac{X}{\tanh(X)}\sqrt{\frac{s}{4s_0}}},\nonumber\\ 
w_z^{ss}(k_x,s)=\frac{Z_0 c X}{a\sinh(2X)}e^{-\frac{X}{\coth(X)}\sqrt{\frac{s}{4s_0}}}. 
\end{align}
These functions have the same two-term Taylor series expansion as the original equations.

Eqs.~(\ref{Eq06}),~(\ref{Eq07}) provide the correct behavior near to the origin, $s=0$, but deviate relatively fast from the exact behavior of the wake function with increase of $s$. For the set of geometric parameters listed in Table~{\ref{Table02}} the distance factor $s_0$  is equal to 0.15 mm. 

To improve the approximation we can follow two ways: to use the next order approximation of the surface impedance $\eta(k)$, Eq.~\ref{Eq05}, or to adjust the distance parameter $s_0$ to fit the results of numerical calculations for the bunch lengths of the interest. The first approach requires an analytical work and "the slivers of insight", see f.e. \cite{StupakovDia}. In this paper we follow the second route.  	
 
   \begin{figure}[htbp]
   	\centering
   	\includegraphics*[height=50mm]{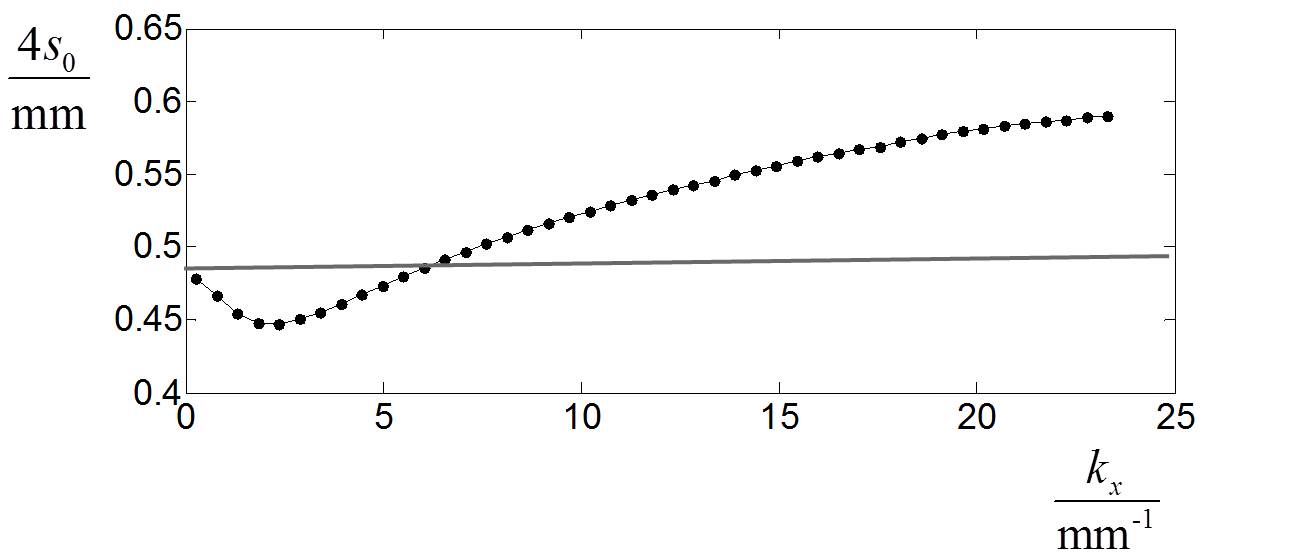}
   	\caption{Parameter $s_0$ as obtained from numerical fit.}\label{Fig05}
   \end{figure}
  
  \begin{figure}[htbp]
  	\centering
  	\includegraphics*[height=70mm]{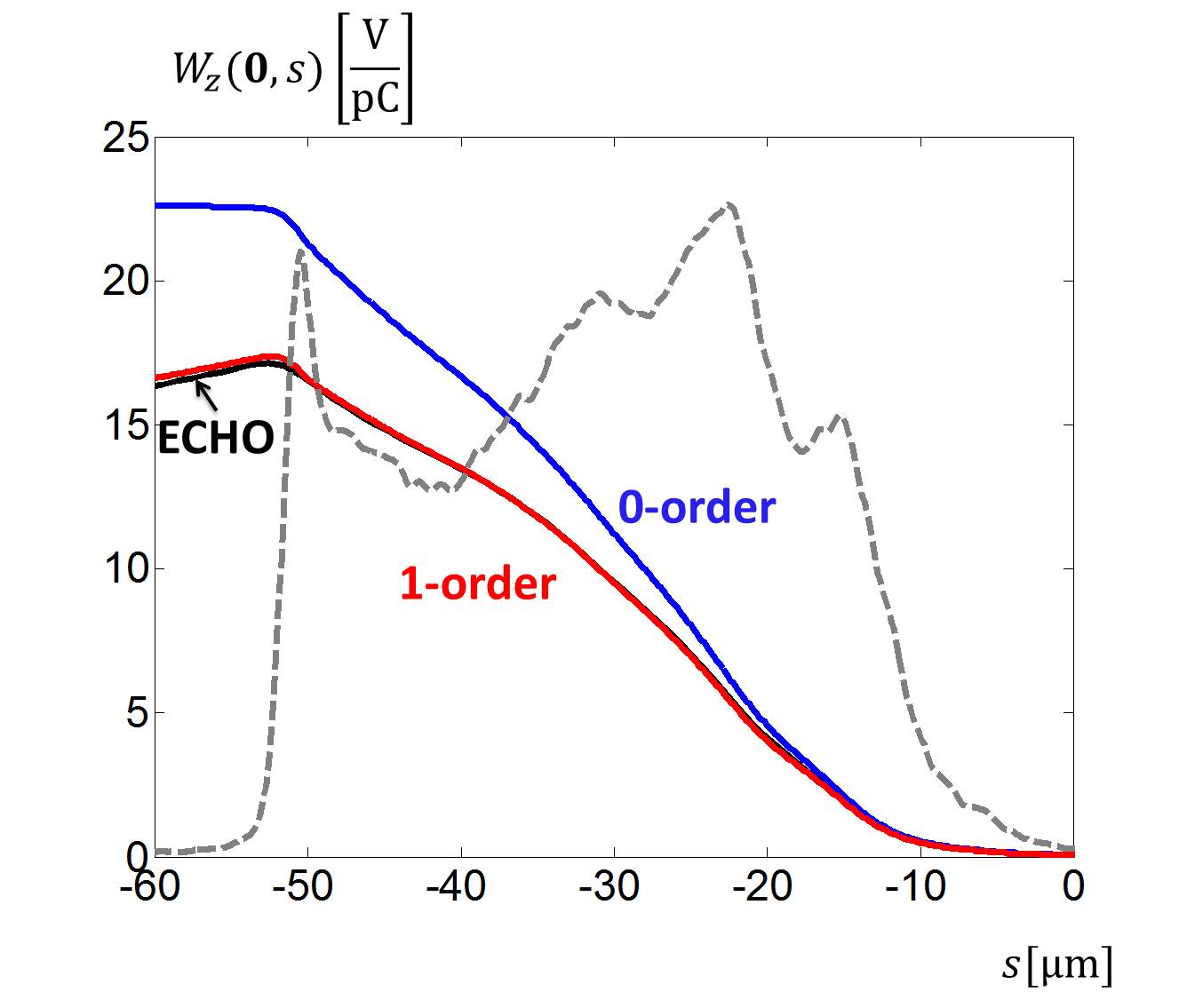}
  	\caption{The longitudinal component of the wake potential on the axis. The results for Eq.~(\ref{Eq07}) (in blue), Eq.~(\ref{Eq04}) (in rot) and ECHO (in black) are shown. The gray dashed line is the current profile.}\label{Fig06}
  \end{figure}
  
  \begin{figure}[htbp]
  	\centering
  	\includegraphics*[height=70mm]{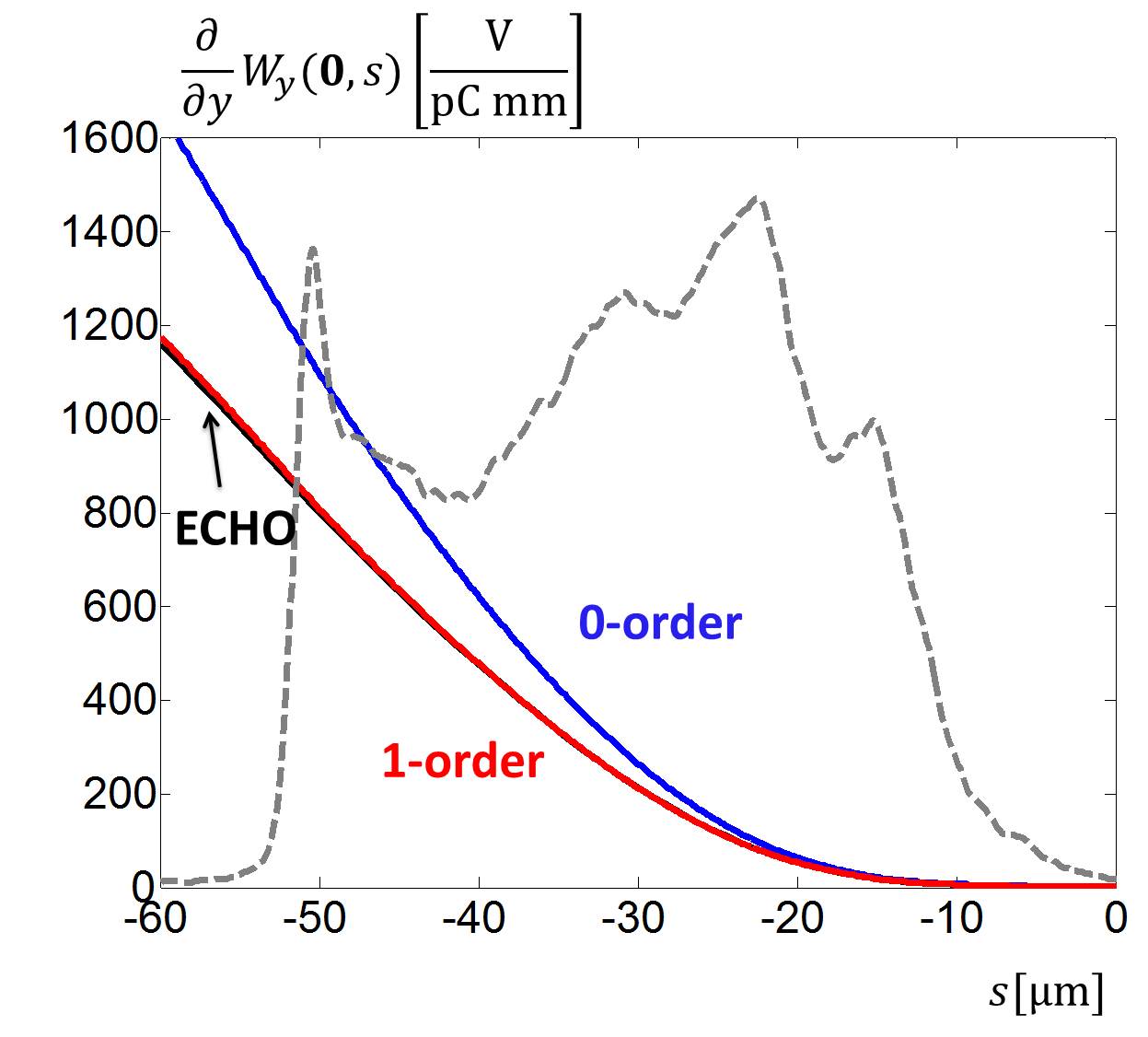}
  	\caption{The quadrupole component of the vertical wake potential on the axis. The results for Eq.~(\ref{Eq07}) (in blue), Eq.~(\ref{Eq04}) (in rot) and ECHO (in black) are shown. The gray dashed line is the current profile.}\label{Fig07}
  \end{figure}
  
  \begin{figure}[htbp]
  	\centering
  	\includegraphics*[height=70mm]{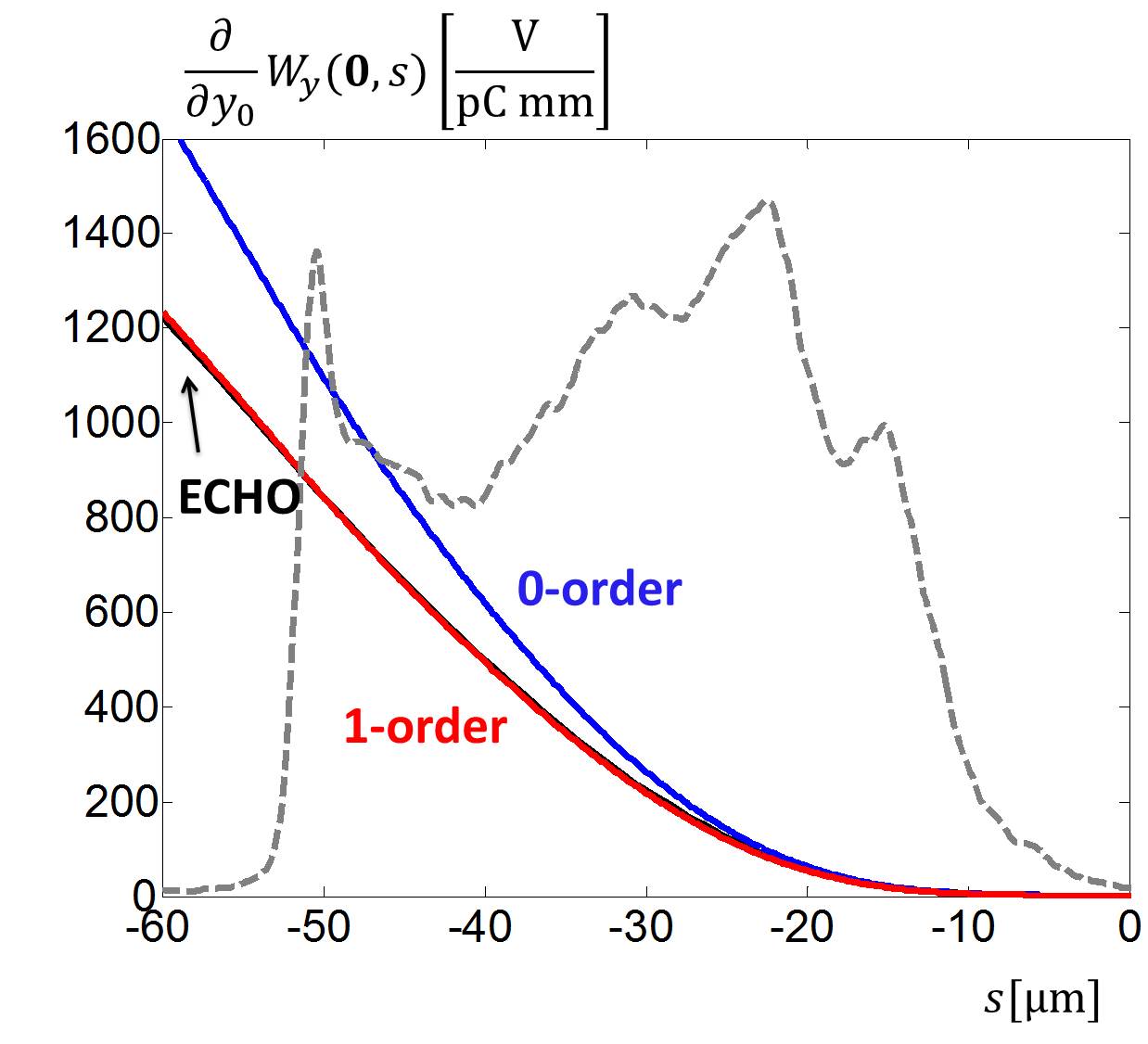}
  	\caption{The "dipole" component of the vertical wake potential on the axis. The results for Eq.~(\ref{Eq07}) (in blue), Eq.~(\ref{Eq04}) (in rot) and ECHO (in black) are shown. The gray dashed line is the current profile.}\label{Fig08}
  \end{figure}
  
 In order to define the parameter $s_0$ we have done sets of simulations with code ECHO~\cite{Zag2015}. The purpose of each set of simulations was to find the "steady-state" modal wake potentials for Gaussian bunches with root-mean-square (RMS) bunch lengthes from 50 $\mu$m to 1 $\mu$m. The values of $s_0$ found from the numerical fitting for each mode are shown by black points in Fig.~{\ref{Fig05}}. The final value for all modes was chosen to be $s_0$=0.12mm (the gray horizontal line in the plot). We note that the same value for $s_0$ can be found from the simple equation
 \begin{equation}\label{Eq08}
 s_0=0.41\frac{a^{1.8} t^{1.6}}{p^{2.4}},
 \end{equation}
 obtained earlier in \cite{Bane1998} for the chain of pillboxes. Eq.(\ref{Eq08}) was obtained in \cite{Bane1998} not analytically but from a numerical fit.
 
 Finally, it follows from Eqs.(\ref{Eq01}),~(\ref{Eq02}), that the longitudinal wake function for arbitrary offsets of the source and the witness particles can be written as
 \begin{align}\label{Eq09}
 &w_z(x_0,y_0,x,y,s)=\frac{1}{w}\sum\limits_{m=1}^\infty 
 w_z(y_0,y,k_{x,m},s)\sin(k_{x,m} x_0)\sin(k_{x,m} x),\\ 
 &w_z(y_0,y,k_x,s)=w_z^{cc}(k_x,k)\cosh(k_x y_0)\cosh(k_x y)+
 w_z^{ss}(k_x,s)\sinh(k_x y_0)\sinh(k_x y),\nonumber
 \end{align}
where $x_0,y_0$ are transverse coordinates of the source particle,  $x,y$ are transverse coordinates of the witness particle, $s=z_0-z$ is the distance in the longitudinal coordinate between the particles. 
 
With direct numerical solution of Maxwell's equation we have checked that for different appertures $a$ between 0.5 and 1.5 mm Eqs.~(\ref{Eq06})-(\ref{Eq09}) provide an accurate approximation of the "steady-state" longitudinal wakefields for arbitrary offsetts of the particles.

The transverse components of the wake function can be derived with help of Panofsky-Wentzel theorem~\cite{Zotter}
 \begin{align}\label{Eq10}
 w_x(x_0,y_0,x,y,s)=-\int_{-\infty}^{s}\frac{\partial  w_\parallel(x_0,y_0,x,y,s')}{\partial x}ds',\nonumber\\
 w_y(x_0,y_0,x,y,s)=-\int_{-\infty}^{s}\frac{\partial  w_\parallel(x_0,y_0,x,y,s')}{\partial y}ds'.
 \end{align}
 
 The wake potential for arbitrary bunch with charge density $\rho(x,y,z)$ can be found by convolution
  \begin{equation}\label{Eq11}
  \vec{W}(x,y,s)=\frac{1}{Q}\int \rho(x_0,y_0,s') \vec{w}(x_0,y_0,x,y,s-s') dx_0 dy_0 ds',
  \end{equation}
where $Q$ is the total bunch charge.  
  
In order to confirm the accuracy of the introduced wake function,~Eqs.~(\ref{Eq06})-(\ref{Eq09}),  we have done in code ECHO the direct calculation of the wake potential for a "pencil" bunch with the same longitudinal current profile as obtained from "start-to-end" simulations, see Fig.\ref{Fig03}. The results of this calculation are shown in Figs.~\ref{Fig06}-\ref{Fig08}. Here we plot the results for Eq.~(\ref{Eq07}) (blue curve) and compare with Eq.~(\ref{Eq04})~(rot curve) and with the direct numerical solution (black curve). We see that the introduced "first-order" approximation agrees accurately with the numerically obtained wake potential for the bunch length of the interest. The gray dashed lines in these plots show the bunch current profile.

Finally let us discuss how we represent the wake function in a tracking program. We consider the structure aligned along $z$-axis and with corrugations in $y$-direction, see Fig.~\ref{Fig04}. The bunch moves near the symmetry axis of the structure $(x,y)=(w,0)$. 

In order to take into account the impact of the corrugated structure on the beam we present the longitudinal wake function through the second order Taylor expansion
\begin{equation}\label{Eq12}
w_z(\vec{r},s)=w_z(\vec{r_a},s)+<\nabla w_z(\vec{r_a},s),\Delta \vec{r}>+\frac{1}{2}<\nabla^2 w_z(\vec{r_a},s) \Delta \vec{r},\Delta \vec{r}>+O(\Delta \vec{r}^3),
\end{equation}
where we have incorporated in one vector the transverse coordinates of the source and the witness particles, $\vec{r}=(x_0,y_0,x,y)^T$,
$\vec{r_a}=(w,0,w,0)^T$, $\Delta \vec{r}=\vec{r}-\vec{r_a}$, and $s$ is a distance between these particles. 

For arbitrary geometry without any symmetry the Hessian matrix $\nabla^2 w_z(\vec{r_a},s)$ contains 8 different elements:
\begin{align}\label{Eq13}
\nabla^2 w_z(\vec{r_a},s)&=
\begin{pmatrix}
h_{11}&h_{12}&h_{13}&h_{14}\\
h_{12}&-h_{11}&h_{23}&h_{24}\\
h_{13}&h_{23}&h_{33}&h_{34}\\
h_{14}&h_{24}&h_{34}&-h_{33}
\end{pmatrix},
\end{align}
where we have taken into account the harmonicy of the wake function in coordinates of the source and the witness particles. 

Hence in general case we use 13 one-dimensional functions to represent the longitudinal component of the wake function for arbitrary offsets of the source and the wittness particles near to the reference axis. For each of this coefficients we use the representation~\cite{zag09}
\begin{equation}\label{Eq14}
 h(s)=w_0(s)+\frac{1}{C}+R c \delta(s)+c\frac{\partial}{\partial s} \left(L c \delta(s)+w_1(s)\right),
\end{equation} 
where $w_0, w_1 $ are non-singular functions, which can be tabulated easily.

The corrugated structure has two planes of symmetry and the wake potential is symmetric in wittness and source coordinates.  Hence near the symmetry axis,  $(x,y)=(w,0)$, we need only 4 one-dimensional functions and Eq.~(\ref{Eq12}) reduces to a very simple one
\begin{align}\label{Eq15}
w_z(x_0,y_0,x,y,s)=&w_z(\vec{r_a},s)+\frac{h_{11}(s)}{2}
\left((x_0-w)^2-y_0^2+(x-w)^2-y^2\right)+\nonumber\\
&+h_{13}(s) (x_0-w)(x-w)+h_{24}(s) y_0 y,\\
h_{11}(s)=\frac{\partial^2 w_z}{\partial x^2}(\vec{r_a},s),\quad
&h_{13}(s)=\frac{\partial^2 w_z}{\partial x_0\partial x}(\vec{r_a},s),\quad
h_{24}(s)=\frac{\partial^2 w_z}{\partial y_0\partial y}(\vec{r_a},s)\nonumber.
\end{align}
where $\vec{r_a}=(w,0,w,0)^T$. 
The transverse components of the corrugated structure wake function can be written as
\begin{align}\label{Eq16}
w_x(x_0,y_0,x,y,s)=-\int_{-\infty}^{s}
\frac{\partial^2 w_z}{\partial x^2}(\vec{r_a},s)ds' (x-w)-\int_{-\infty}^{s}
\frac{\partial^2 w_z}{\partial x_0 \partial x}(\vec{r_a},s)ds' (x_0-w),\\
w_y(x_0,y_0,x,y,s)=\int_{-\infty}^{s}
\frac{\partial^2 w_z}{\partial x^2}(\vec{r_a},s)ds' y+\int_{-\infty}^{s}
\frac{\partial^2 w_z}{\partial y_0 \partial y}(\vec{r_a},s)ds' y_0.\label{Eq17}
\end{align}

 %
 \section{Beam dynamics in flat corrugated structure }\label{sec:4}
 
 \begin{figure}[htbp]
 	\centering
 	\includegraphics*[height=30mm]{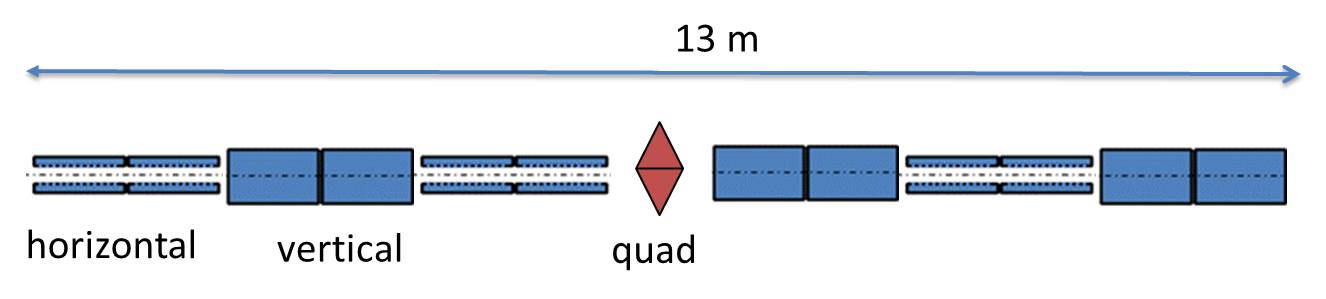}
 	\caption{Layout of the corrugated structure insertion. The modules have length of 2 meters and are alternatively oriented.}\label{Fig09}
 \end{figure}
 
 In this section we present results of particle tracking through the setup shown in Fig.~\ref{Fig09}. We suggest that the corrugated structure insertion will be placed at the end of the accelerator. Here the magnet lattice consists of equally spaced focusing and defocusing quadrupoles ("FODO" lattice) with period of 30 meters.  The setup consists of 6 corrugated structures with one quadrupole in the middle. In order to compensate the defocussing effect of the "quadrupole" component of the wake, shown in Fig.~\ref{Fig07}, the neighbor corrugated structure modules are rotated by 90 degree.
  
  If the wake potential $\vec{W}(x,y,s)$ of the bunch with charge density  $\rho(x,y,z)$ is known then the change in the momentum for particle with charge $q$ in bunch of total charge $Q$  can be found as
  \begin{equation}\label{Eq18}
  \Delta \vec{p}(x,y,s)=\frac{q}{c}\vec{W}(x,y,s)=\frac{q}{cQ}\int \rho(x_0,y_0,s') \vec{w}(x_0,y_0,x,y,s-s') dx_0 dy_0 ds',
  \end{equation}
  where $c$ is the velocity of light. 
  
  In order to model the beam dynamics in the presence of the wakefields we have used the open source code OCELOT \cite{Agapov2015}. We have developed and tested the wake field module. The implementation follows closely the approach described in~\cite{zag09},~\cite{Dohlus}.
  The wake field impact on the beam is included as series of kicks. Note, that the code ASTRA can be used for such kind of simulations as well~\cite{Dohlus}. We have done a comparison between ASTRA and OCELOT. At the initial tests the results of ASTRA for the strongly chirped bunch  were		 unphysical. This bug is now fixed by ASTRA's developers and the results from both codes agree.
  
  For testing of the numerical implementation we have compared the results from code OCELOT with the analytical estimations presented in~\cite{Bane2016a}. It is shown in~\cite{Bane2016a} that for "zeroth"-order wake function and a rectangular bunch profile of length $l$ the emittance growth $\epsilon/\epsilon_0$ , the energy loss $\Delta E(s)$ and the energy spread $\sigma_E(s)$ along the bunch can be estimated as
  \begin{align}\label{Eq19}
  &\frac{\epsilon}{\epsilon_0}=\sqrt{1+\left(\frac{\pi^3 Z_0 c e Q \beta L l}{384 \sqrt{5} a^4 E} \right)^2},\qquad
  \Delta E(s)=\frac{\pi Z_0 c e Q L s}{16 a^2 l},\nonumber\\
  &\sigma_E(s)=\frac{\sqrt{2}\pi^3 Z_0 c e Q L s}{256 a^4 l} \sqrt{\sigma_x^4+\sigma_y^4}.
  \end{align}
  The parameters used in these equations can be found in Table~\ref{Table03}. The beam parameters of the "ideal" rectangular beam are chosen as average parameters of the beam obtained with the "start-to-end" simulations described in Section~\ref{sec:2}. The values of optical $\beta$-functions correspond to the design values of the "FODO" cell at the European XFEL lattice. We start the tracking at the position where the optical $\alpha_x,\alpha_y$ functions are equal in absolute value.
   
  \begin{table}[htbp]
  	\centering
  	\caption{"Ideal" beam parameters.}
  	\label{Table03}
  	\begin{tabular}{lcc}
  		{\bf Parameter name}					&{\bf Value}	&{\bf Unit}\\
  		Emittance,~$\epsilon_{0,x}/\epsilon_{0,y}$	   	& 0.64/1.09 	& $\mu$m \\
  		RMS size,~$\sigma_x/\sigma_y$  			& 30.9/16.8		& $\mu$m \\
  		Beta function,~$\beta_x/\beta_y$  		& 22.6 			& m \\
  		Alpha function,~$\alpha_x/\alpha_y$  	& -1.43/1.43	& \\
  		Energy,~$E$ 							& 14			& GeV \\
  		Length,~$L$  							& 40			& $\mu$m \\
  		Charge,~$Q$   							& 0.5 			& nC \\
  	\end{tabular}
  \end{table}
  
  In this example we use a "cold" beam with zero energy spread and apply only one kick in OCELOT without particle tracking. The kick corresponds to the structure of length 2 m.  As we see from Table~\ref{Table04} the numerical and analytical results  for "zeroth"-order approximation agree. But they overestimate considerably the real wake field effects. The wake described accurately with "first"-order approximation is given in the last column of~Table~\ref{Table04}. The energy loss in tail introduced by one module is only 35 MeV. Hence in order to introduce 1.5\% of additional energy loss at 14 GeV we need at least 6 modules. Another possibility is to reduce the aperture $2a$ between the corrugated plates. In the following we will study only the case with the nominal aperture $2a=1.4$ mm.
     
   \begin{table}[htbp]
   	\centering
   	\caption{Comparison of the numerical results with the analytical estimations.}
   	\label{Table04}
   	\begin{tabular}{lcccc}
   		{\bf Parameter name }	&{\bf Analyt.\cite{Bane2016a}}	&{\bf "0th"-order} 	&{\bf "1st"-order}& {\bf Units}\\
   		Emittance growth		& 1.484 		& 1.479					& 1.29& \\
   		Energy spread at tail	& 80.2			& 81.0					& 56& keV\\
   		Energy loss at tail 	& 45.3 			& 45.0					& 35&MeV\\
   		\end{tabular}
   \end{table}
  
  In the following we will consider not only the beam from the "start-to-end" simulations shown in Fig~\ref{Fig03}, but additionally the "ideal" beam with a rectangular longitudinal profile, without energy chirp and with constant slice parameters listed in Table~\ref{Table03}. The slice energy spread is taken to be 500 keV.

  \begin{figure}[htbp]
  	\centering
  	\includegraphics*[height=60mm]{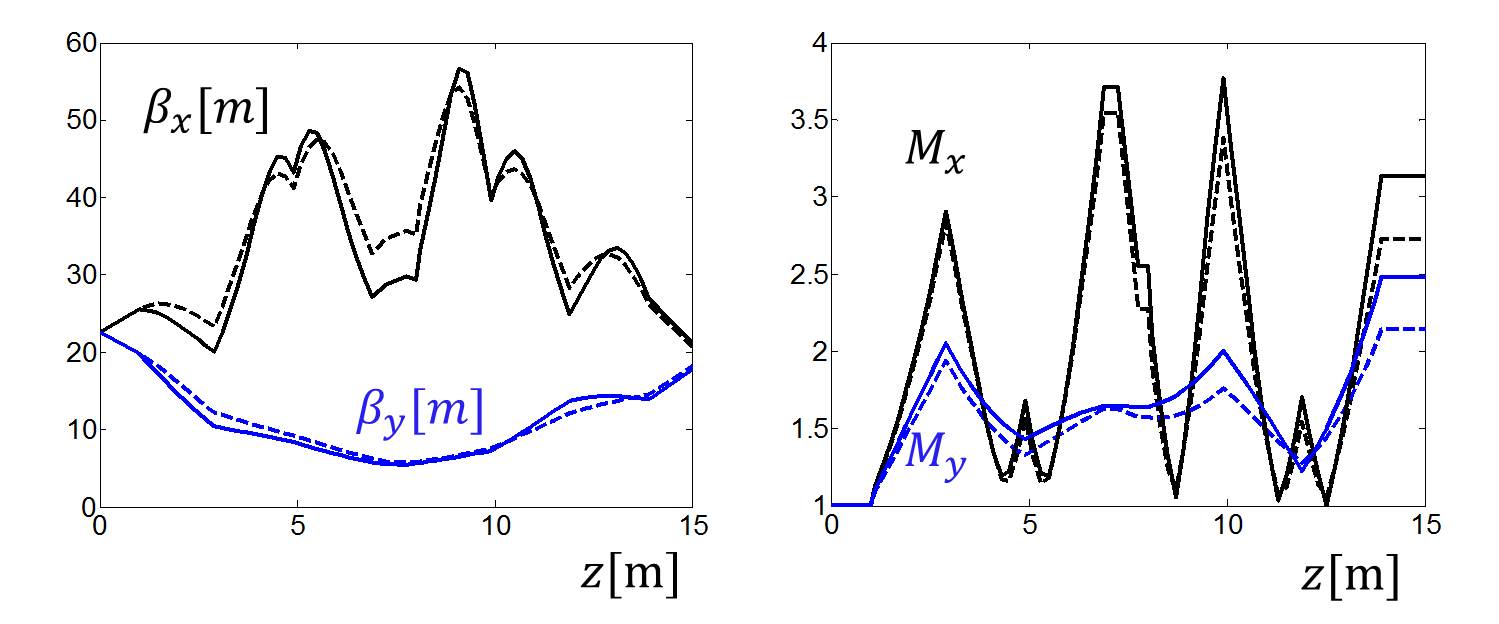}
  	\caption{Beam $\beta$-functions (in the left plot) and beta mismatch parameters~\cite{Sands} (in the right plot) along the corrugated strucure insertion. The solid lines correspond to the beam from "start-to-end" simulations, the dashed lines describe the results for the "ideal" rectangular beam. }\label{Fig10}
  \end{figure}
  
     \begin{figure}[htbp]
     	\centering
     	\includegraphics*[height=70mm]{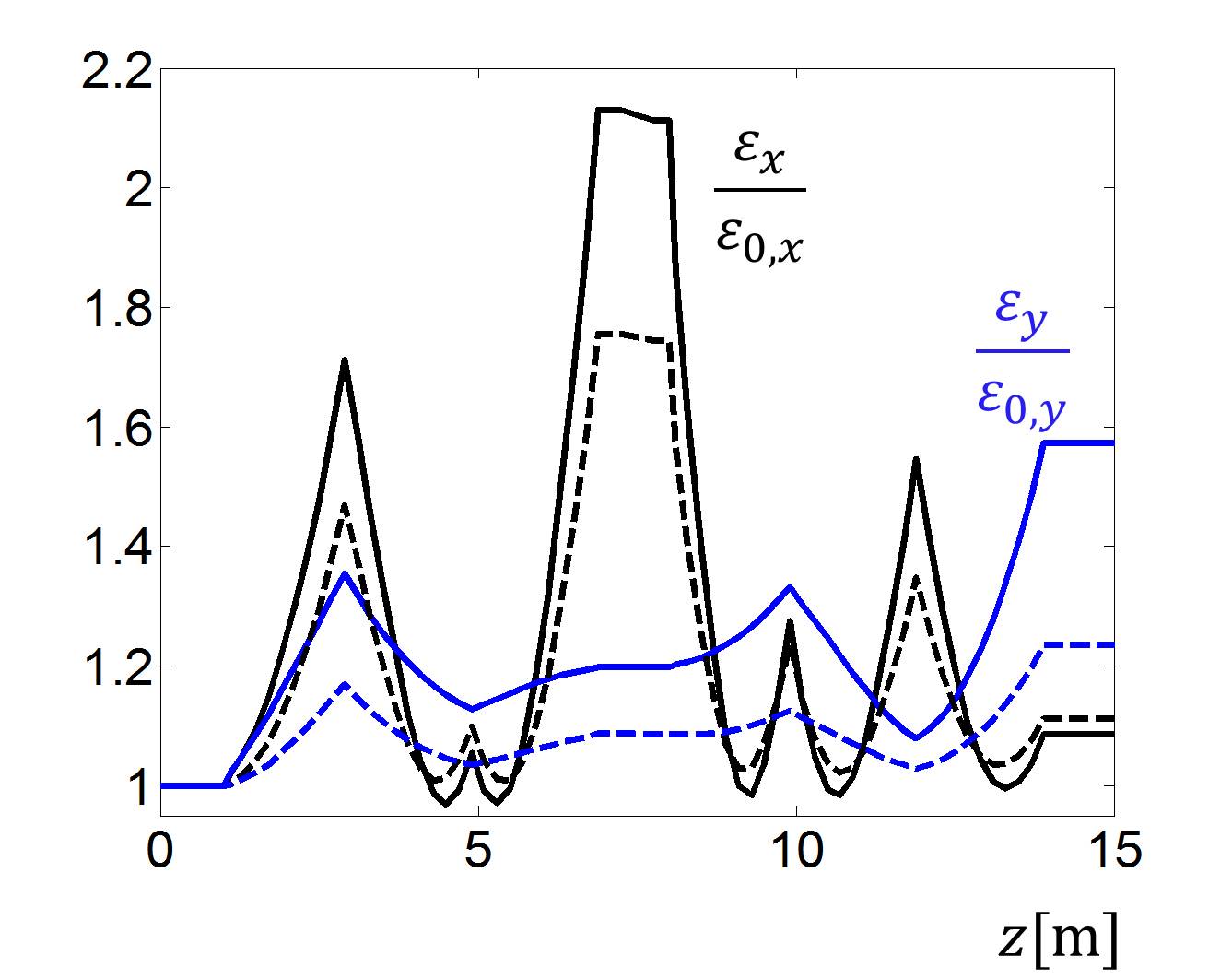}
     	\caption{Projected emittance growth along the corrugated strucures insertion. The solid lines correspond to the beam fron "start-to-end" simulations, the dashed lines describe the results for the "ideal" rectangular beam. }\label{Fig11}
     \end{figure}

  The tracking is done through the setup of Fig.~\ref{Fig09} along distance of 15 meters. We start at the position of 1 meter before the first module and end at 1 meter after the last one. The beam is perfectly matched to the optics with quadrupole strength $k=0.1905$. Due to the wakefields the beam will be mismatched from the design optics. In order to reduce the mismatch we have increased the strength of the quadrupole between the corrugated structure modules to $k=0.215$. We have done no further optimization of the optics. The optical $\beta$-functions of the beam and the beta mismatch parameters~\cite{Sands} along the insertion are shown in Fig.~\ref{Fig10}. The beta mismatch parameter $M$ in each projected transverse plane ($xx'$ or $yy'$) is defined as
     \begin{align}\label{Eq20}
   M=\frac{1}{2}\left(\tilde{\beta}_e+\tilde{\gamma}_e+\sqrt{\left(\tilde{\beta}_e+\tilde{\gamma}_e\right)^2-4}\right),\\
   \tilde{\gamma}_e=\frac{1+\tilde{\alpha}_e^2}{\tilde{\beta}_e},\quad
   \tilde{\alpha}_e=\alpha_e-\alpha\tilde{\beta}_e,\quad
   \tilde{\beta}_e=\frac{\beta_e}{\beta},\nonumber
   \end{align}
   where $\alpha, \beta$ correspond to the design optics, $\alpha_e, \beta_e$ are obtained from the tracked beam. in Fig.~\ref{Fig10} a considerable mismatch in the beam optics at the end of the insertion can be seen. We think that this mismatch can be corrected with  some quadrupoles after the insertion.
   
   In Fig.~\ref{Fig11} we show the projected emittance growth along the insertion. A compensation of the projected emittance growth due to the alternative orientation of the corrugated structure modules can be seen. However at the end we see about 60\% growth in the vertical emittance. This increase in the projected emittance cannot be corrected and it  will result in the slice mismatch along the beam at the undulator, see Fig.~\ref{Fig14}. Impossibility to match all slices along the beam will reduce the SASE radiarion power, see Section~\ref{sec:5}.
   
  \begin{figure}[htbp]
  	\centering
  	\includegraphics*[height=60mm]{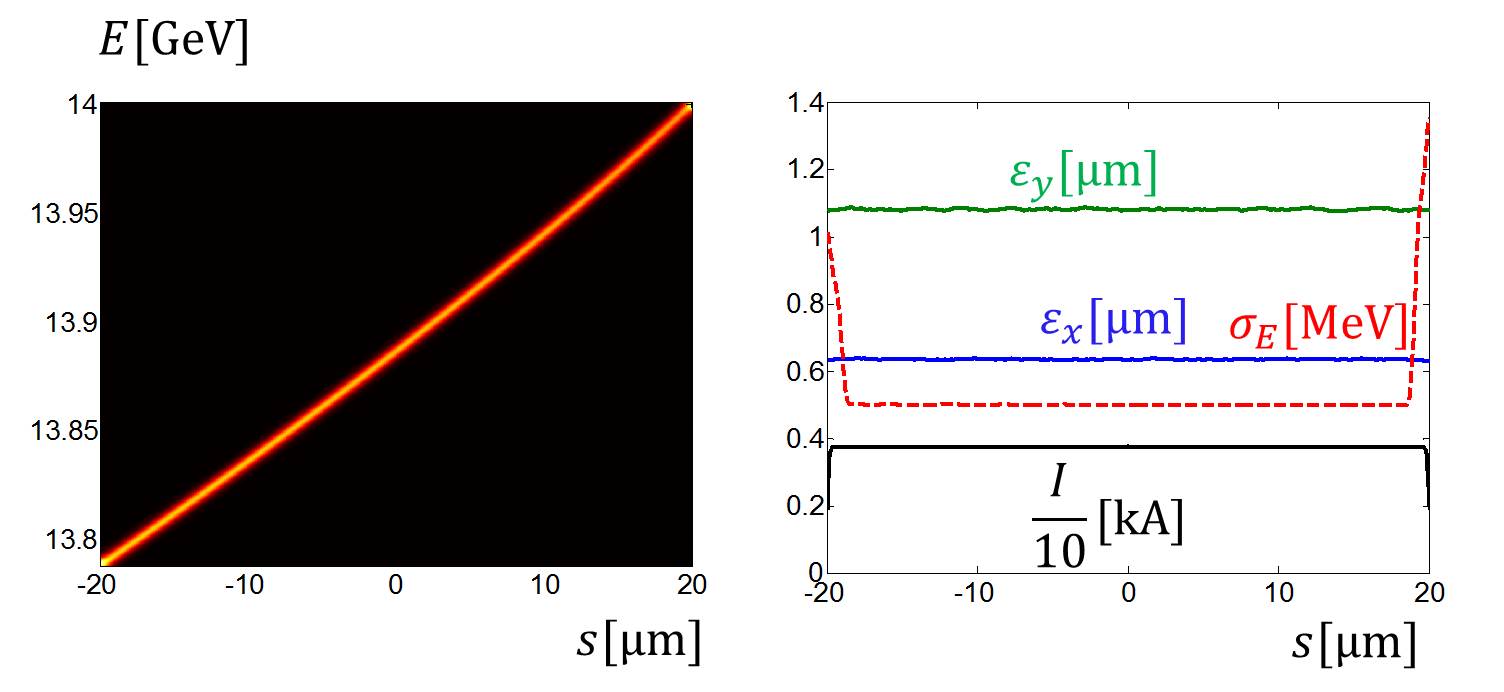}
  	\caption{The longitudinal phase space (in the left plot), the current profile and the slice parameters (in the right plot) for the "ideal" rectangular beam.}\label{Fig12}
  \end{figure}
   \begin{figure}[htbp]
   	\centering
   	\includegraphics*[height=60mm]{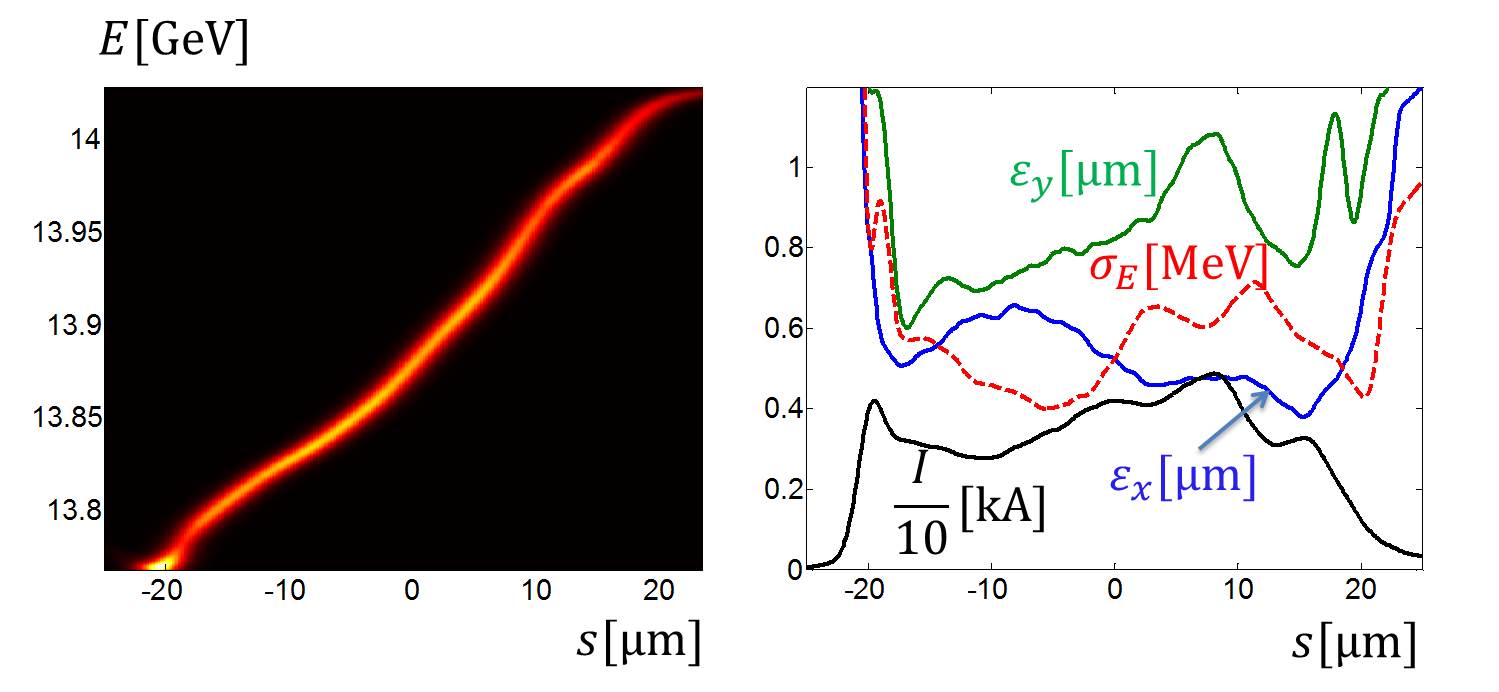}
   	\caption{The longitudinal phase space (in the left plot), the current profile and the slice parameters (in the right plot) for the beam from "start-to-end" simulations.}\label{Fig13}
   \end{figure} 
  
  In Fig.~\ref{Fig12} and Fig.~\ref{Fig13} the longitudinal phase space (in the left plots) and the slice parameters of the beams (in the right plots) after the corrugated structure insertion are shown. There is almost no change in the electron beam slice parameters. It can be seen that, due to the alternative orientation of the modules and the large initial energy spread in the electron bunch, the increase of the final slice energy spread at the bunch tail is negligible.
      
   \begin{table}
     	\caption{Impact of the corrugated structure insertion on the beam parameters.}
     	\label{Table05}
     	\begin{tabular}{lccc}
     		{\bf Parameter name}	&{\bf "ideal" beam}	&{\bf "s2e" beam} 	&{\bf Units}\\
			 Initial/final projected $x$-emittance  			
									& 0.64/0.70			& 		0.72/0.77		& $\mu$m\\
     		 Initial/final projected $y$-emittance,   			
						     		 & 1.09/1.33 		& 		1.18/1.82		& $\mu$m\\
     		 Initial/final energy loss at tail	
						     		 & 0/212 			& 		50/255			& MeV\\
     	\end{tabular}
     \end{table}
 
 The changes in the main parameters of the beam after the insertion are summarized in Table~\ref{Table05}. We see that the relative energy difference between the tail and the head of the beam from "start-to-end" simulations reached now 1.85\% and we can hope that the SASE radiation bandwidth will exceed 3\%. 
  
 Finally, let us considered another important issue. The energy lost by the beam will go to Joule heating in the metal walls, to the energy in the THz pulse that leaves the end of the insertion and to the energy leaking out of the sides of the structure. A detailed analysis of the situation can be found in~\cite{Bane2016c}. Here we will do a suggestion that the energy of the wake field goes completely into the walls. To obtain the power absorbed into the walls, we take
 \begin{align}\label{Eq19}
	 P=k_{loss}Q^2 N_b f_{rep},  
  \nonumber\
 \end{align}
 with $k_{loss}$ the loss factor, $Q$ the beam charge, $N_b$ the number of bunches in one train and $f_{rep}$ the train repetition rate. For the bunch from the "start-to-end" simulations the loss factor is equal to 18.6 MV/nC per meter. For the nominal number of bunches, $N_b=2700$, and the highest train repetition rate, $f_{rep}=30$ Hz, the power to be absorbed in the walls is equal to 377 W per meter. Hence the Joule heating of the corrugated structure modules by the beam’s wake fields is significant and a cooling system may be required. 
 
 In addition, the collimator system of the European XFEL~\cite{Balandin} is not designed for the small aperture of the corrugated structure. Halo particles  or even mis-steered bunches might hit the structure directly if no countermeasures will be taken.

 %
 \section{Broadband SASE radiation in undulator section}\label{sec:5}
  
  In this section we consider the properties of SASE radiation in the undulator lines SASE1/SASE2 of the European XFEL. The parameters of the undulators and the expected radiation wavelength for different electron beam energies are listed in Table~\ref{Table06}~\cite{Pfl2015}. With the nominal scenario of the bunch compression and the transport the electron bunch with charge 0.5 nC and energy 14 GeV will have only about several MeV of energy spread and will produce the SASE radiation with "full-width-half-maximum" (FWHM) bandwidth 0.2\% at photon energy of 4.96 keV~\cite{Yurkov2011}.
  With the scenario described in Section~\ref{sec:2} and with the corrugated structure insertion of Section~\ref{sec:4} the energy difference between the tail and the head is encreased to 255 MeV. However we have undesirable increase in the projected electron bunch emittance and we need a numerical modelling of the free electron physics in the undulator in order to estimate the properties of the expected SASE radiation.
  
  In our FEL simulations we use code ALICE~\cite{Zag2012}. The electron beam is described with help of slice parameters. Each beam slice with coordinate $s$ has the following parameters: mean energy $E(s)$, mean positions $x(s)/y(s)$, mean momentums $p_x(s)/p_y(s)$, RMS energy spread $\sigma_E(s)$ , emittances $\epsilon_x(s)/\epsilon_y(s)$, beta functions $\beta_x(s)/\beta_y(s)$, gamma functions $\gamma_x(s)/\gamma_y(s)$, current $I(s)$. 
  The parameters of the undulator section are listed in Table~\ref{Table07}.

  \begin{table}
  	\caption{The XFEL undulator lines~\cite{Pfl2015}.}
  	\label{Table06}
  	\begin{tabular}{lccc}
  		{\bf Parameter name}	&{\bf SASE1/SASE2}	&{\bf SASE3} 	&{\bf Units}\\
  		undulator wavelength  	& 40				& 		68		& mm\\
  		K-range  				& 3.9-1.65			& 		9.3-4	& \\
  		Radiation wavelength at 17.5 GeV& 0.147-0.040&1.22-0.27 	&nm \\
  		Radiation wavelength at 14.0 GeV& 0.230-0.063&1.90-0.42 	&nm \\
  		Radiation wavelength at 17.5 GeV& 0.625-0.171&5.17-1.15 	&nm \\
  		Active undulator length & 175 				& 105	 		&m \\
  		\end{tabular}
  \end{table}

 \begin{table}
 	\caption{Parameters for FEL simulations.}
 	\label{Table07}
 	\begin{tabular}{lcc}
 		{\bf Parameter name}	&{\bf Value}	&{\bf Units}\\
 		undulator wavelength  	& 40			& mm\\
 		K averaged  			& 	2.76		& 	\\
 		Radiation wavelength	& 0.23			& nm \\
 		averaged $\beta$-function& 	16			& m	\\
 	\end{tabular}
 \end{table}

   \begin{figure}[htbp]
   	\centering
   	\includegraphics*[height=60mm]{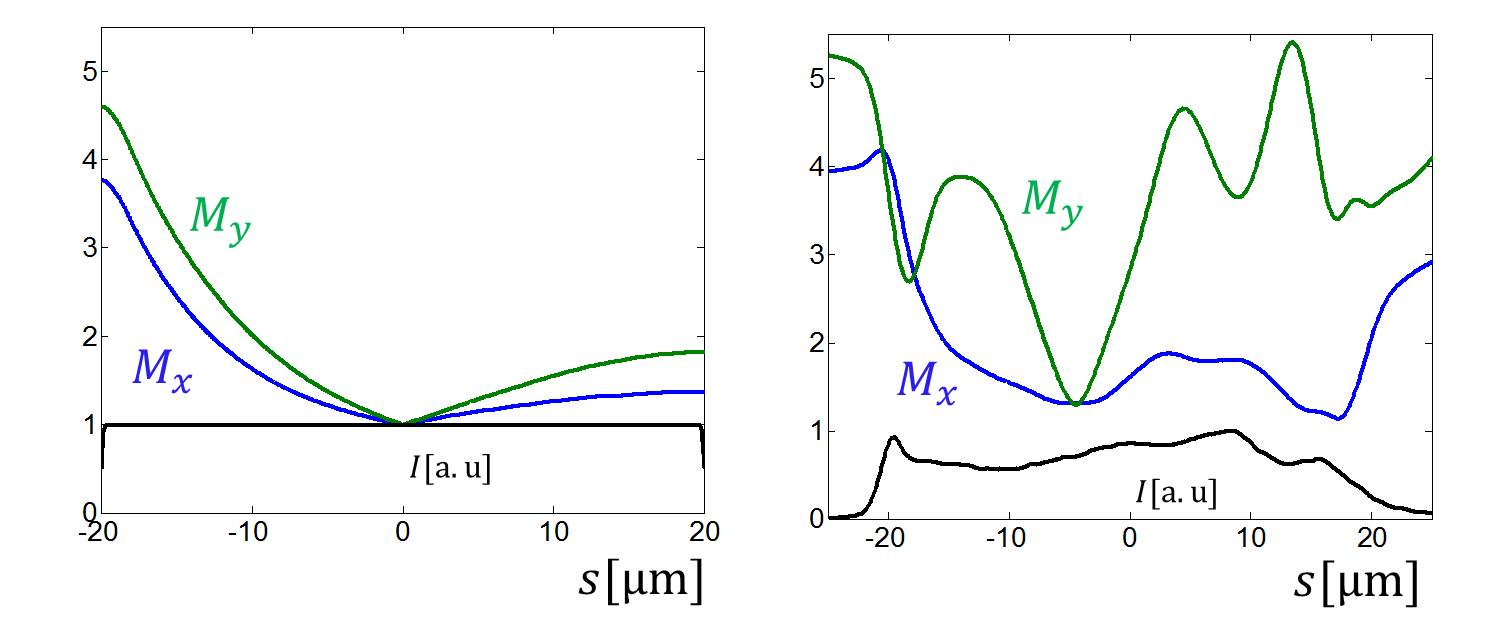}
   	\caption{The mismatch parameters for the "ideal" beam (left plot) and the beam from "start-to-end" simulations (right plot).}\label{Fig14}
   \end{figure} 
  
 \begin{figure}[htbp]
 	\centering
 	\includegraphics*[height=60mm]{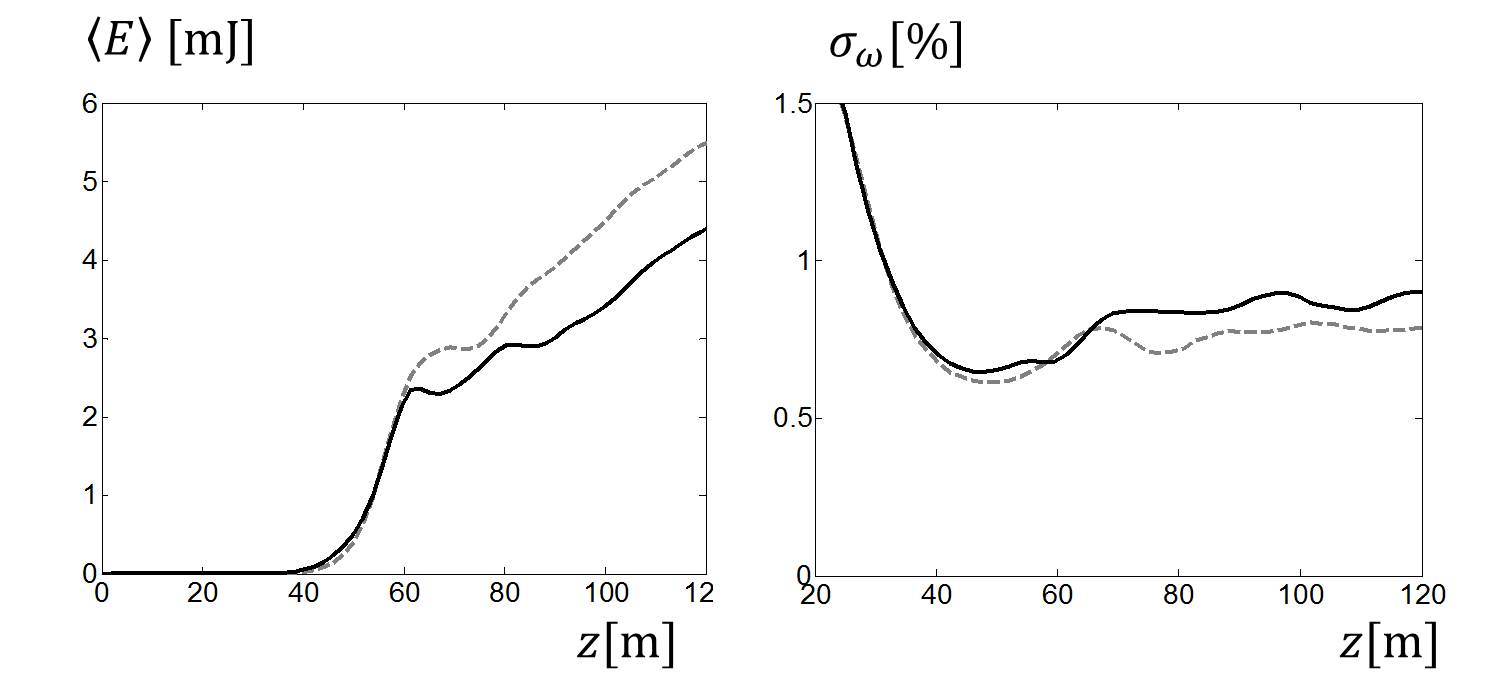}
 	\caption{The radiation energy (left polt) and the RMS bandwidth (right plot). The solid line corresponds to the beam from "start-to-end" simulation, the dashed line presents the results for the "ideal" beam.}\label{Fig15}
 \end{figure} 
 
  \begin{figure}[htbp]
  	\centering
  	\includegraphics*[height=60mm]{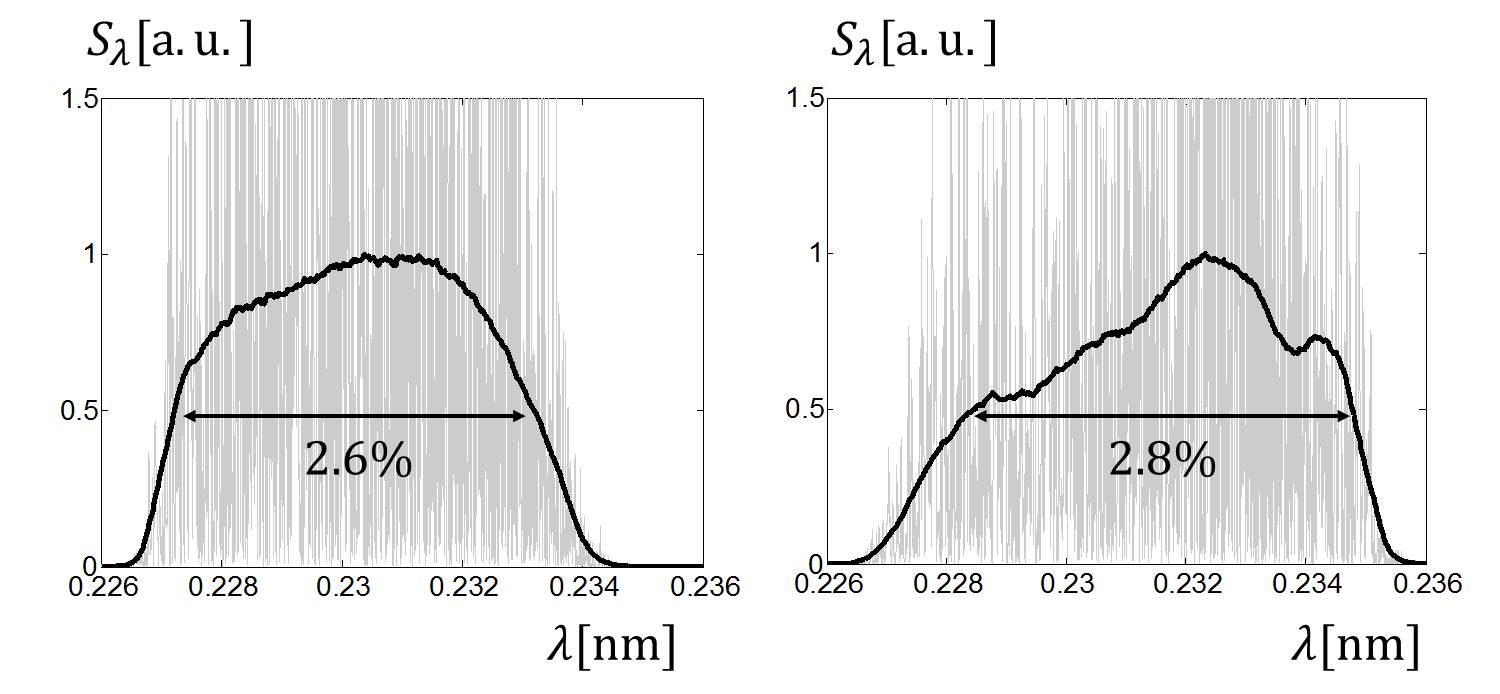}
  	\caption{The radiation spectrum at z=115 m for the "ideal" beam (left plot) and the beam from "start-to-end" simulations (right plot).}\label{Fig16}
  \end{figure} 
  
	\begin{figure}[htbp]
	\centering
	\includegraphics*[height=60mm]{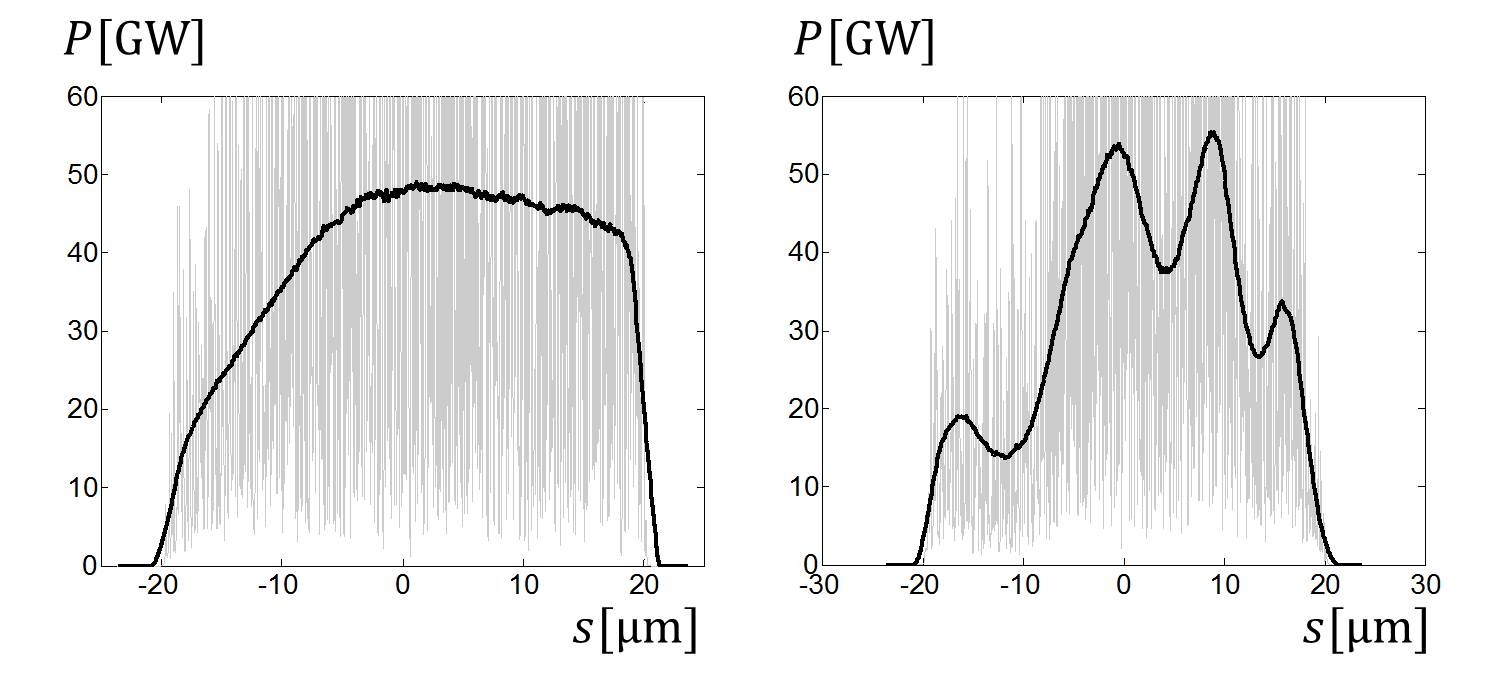}
	\caption{The radiation power at z=115 m for the "ideal" beam (left plot) and the beam from "start-to-end" simulations (right plot).}\label{Fig17}
	\end{figure} 

 In Fig.~\ref{Fig14} we show the beta mismatch parameters~\cite{Sands} along the beam matched to the undulator entrance. We use the same Eq.~(\ref{Eq20}), but $\alpha, \beta$ correspond now to the design optics at the undulator entrance, $\alpha_e(s), \beta_e(s),$ are the slice values along the beam. The "ideal" beam can be matched relatively well even at the tail. The beam from the "start-to-end" simulations has a larger mismatch at the head and at the tail. Indeed the radiation properties depend strongly on the way of the beam matching. In our first try we have matched the beam perfectly at the middle point. But in this case the mismatch at the tail was very large ($M_y>10$)  and the radiation from the tail was suppressed heavily. The case shown in Fig.~\ref{Fig14} is obtained when we reduce the mismatch parameters along the whole bunch. 
 
 The active length of the undulator is 175 meters. With the electron beam parameters under consideration the saturation of the SASE radiation is reached at the position $z=60$ m. In the left plot in Fig.~\ref{Fig15} we show the averaged energy in the photon pulse along the undulator line. The radiation pulse energy at the saturation is about 3 mJ. We do not use any undulator tapering and do not take into account the undulator wake fields in this study. The tapering could increase the radiation power yet by order of magnitude but the radiation bandwidth can be reduced because of it. The impact of the tapering and the undulator wakefields will be studied in another work. 
 
 The RMS radiation bandwidth along the undulator section is shown in the right plot in Fig.~\ref{Fig15}. It reaches 0.9\% at position $z=120$ m. The radiation spectrum has not Gaussian shape. In Fig.~\ref{Fig16} we show the spectrum at position $z=115$ m. The FWHM radiation bandwidth reaches here 2.8\% for the electron beam from "start-to-end" simulations. The solid lines present the spectrum averaged over many shots. The oscillating gray lines show an one shot spectrum. 
   
 The radiation power along the pulse can be seen in Fig.~\ref{Fig17}. The radiation from the beam tail and head are suppressed partially due to impact of the wake fields on the beam quality in the corrugated structure insertion.The solid lines present the averaging over many shots.
  
 %
\section{Conclusion}
 
 In this paper we have studied a possibility to extend the bandwith of the radiation at the European XFEl with the help of a special compression scenario together with the corrugated structure insertion. We have derived an accurate modal representation of the wake function of corrugated structure  and have applied this fully three dimensional wake function in beam dynamics studies in order to estimate the change of the electron beam properties. In the second part of the study we have done simulations of free electron laser physics with the electron bunches obtained from the beam dynamics simulations. It was shown that at photon energy 5.4 keV such scenario can produce a photon pulse with a 3\% bandwidth, a few microjoule radiation pulse energy, and a few femtosecond pulse duration. The Joule heating of the corrugated structure modules by the beam’s wakefields is significant and a cooling system may be required.
 
 \section{ACKNOWLEDGMENTS}
 
 The authors thank W. Decking,  M. Dohlus and J. Zemella for helpful discussions.

\end{document}